\documentclass[aps,prx,twocolumn,superscriptaddress,showpacs,floatfix,longbibliography]{revtex4-1}
\usepackage{amsmath,amssymb,amsfonts,float,graphics,epsfig,epstopdf,color,verbatim,tabularx,bm,multirow,appendix}
\usepackage[utf8]{inputenc}
\usepackage[T1]{fontenc}
\usepackage{xcolor}

\IfFileExists{newtxtext.sty}
{\usepackage{newtxtext,newtxmath}}
{\IfFileExists{stix.sty}
	{\usepackage{stix}}
	{\IfFileExists{mathptmx.sty}
		{\usepackage{mathptmx}}{} } }

\usepackage{textcomp}
\usepackage{bm}

\IfFileExists{siunitx.sty}{\usepackage{booktabs,siunitx}}{}

\usepackage{color}
\definecolor{LinkColor}{rgb}{0.256,0.439,0.588}
\usepackage{hyperref}
\hypersetup{
	pdfauthor={Yuzhi Liu},
	pdftitle={paper},
	colorlinks=true,
	citecolor=LinkColor,
	linkcolor=LinkColor,
	urlcolor=LinkColor
}

\usepackage{pifont}

\providecommand{\U}[1]{\protect \rule{.1in}{.1in}}

\begin{document}
\title{Designer Monte Carlo Simulation for Gross-Neveu-Yukawa Transition}
\author{Yuzhi Liu}
\affiliation{Beijing National Laboratory for Condensed Matter Physics and Institute of Physics,Chinese Academy of Sciences, Beijing 100190, China}
\affiliation{School of Physical Sciences, University of Chinese Academy of Sciences, Beijing 100190, China}
\author{Wei Wang}
\affiliation{Beijing National Laboratory for Condensed Matter Physics and Institute of Physics,Chinese Academy of Sciences, Beijing 100190, China}
\affiliation{School of Physical Sciences, University of Chinese Academy of Sciences, Beijing 100190, China}
\author{Kai Sun}
\affiliation{Department of Physics, University of Michigan, Ann Arbor, MI 48109, USA}
\author{Zi Yang Meng}
\affiliation{Department of Physics and HKU-UCAS Joint Institute of Theoretical and Computational Physics, The University of Hong Kong, Pokfulam Road, Hong Kong, China}
\affiliation{Beijing National Laboratory for Condensed Matter Physics and Institute of Physics,Chinese Academy of Sciences, Beijing 100190, China}
\affiliation{Songshan Lake Materials Laboratory, Dongguan, Guangdong 523808, China}

\begin{abstract}
In this manuscript, we study quantum criticality of Dirac fermions via large-scale numerical simulations, focusing on the Gross-Neveu-Yukawa (GNY) chiral-Ising quantum critical point with critical bosonic modes coupled with Dirac fermions. We show that finite-size effects at this quantum critical point can be efficiently minimized via model design, which maximizes the ultraviolet cutoff and at the same time places the bare control parameters closer to the nontrivial fixed point to better expose the critical region. Combined with the efficient self-learning quantum Monte Carlo algorithm, which enables non-local update of the bosonic field, we find that moderately-large system size  (up to $16\times 16$) is already sufficient to produce robust
scaling behavior and critical exponents.
The conductance of free Dirac fermions is also calculated and its  frequency dependence is found to be consistent with the scaling behavior predicted by the conformal field theory. 
The methods and model-design principles developed for this study can be generalized to other fermionic QCPs, and thus provide a promising direction for 
controlled studies of strongly-correlated itinerant systems.
\end{abstract}
\date{\today}

\maketitle

\section{Introduction}
\label{sec:i}

Quantum critical points (QCP) in the presence of fermionic fluctuations are among the most intriguing topics in the study of strongly correlated systems, which pave the way towards new paradigms of quantum matter beyond the conventional Fermi-liquid theory of metal and the Landau-Ginzburg-Wilson theory of phase transitions. Due to the strong coupling nature, to understand these fermionic QCPs remains a highly challenging task, and many fundamental questions remain open. For example,  after four decades of intensive studies, quantum criticality in the presence of Fermi surfaces still remains a highly active research subject, from the early effort of the Hertz-Mills-Moriya framework~\cite{Hertz1976,Millis1993,Moriya1985} based on the leading-order approximation to the more recent studies that reveal the crucial and nontrivial contributions of higher order terms~\cite{Abanov2003,Chubukov2004,Chubukov2009,SSLee2009}. Even today, the fate of such fermionic QCPs still remains an important open question and a source for exciting new ideas and insights~\cite{SSLee2009,Metlitski2010a,Metlitski2010b,Schlief2017}.

The situation at the numeric front is equally, if not more, challenging, as the divergent length scale and the fermionic nature of this problem prevents us from obtaining unbiased and reliable numerical solutions at the thermodynamic limit. For example, in contrast to bosonic QCPs, many of which can be efficiently simulated via quantum Monte Carlo (QMC) methodologies~\cite{sandvik2010}, the notorious sign problem makes it highly challenging to access the same level of knowledge in the presence of fermionic fluctuations. Only till very recent, new pathways towards large-scale unbiased simulations of fermoinc QCPs become available, where utilizing designers models, various fermoinc QCPs become accessible to sign-problem-free QMC simulations, such as nematic~\cite{Schattner2016}, ferromagnetic~\cite{Xu2017} and antiferromagnetic QCPs~\cite{Gerlach2017,ZiHongLiuTri2018,ZiHongLiuEMUS2018,ZiHongLiuSqu2018} (See Ref.~\cite{XYXu2019} for a recent review). From these efforts, it becomes possible to obtain accurate and reliable information about the scaling behaviors in the close vicinity of these QCPs, to test and improve our theoretical knowledge about these challenging problems.

In this manuscript, we study fermionic QCPs in strongly-correlated Dirac systems utilizing Gross-Neveu-Yukawa type of models, where interactions between Dirac fermions are mediated by a bosonic field. These QCPs are often refereed to as the Gross-Neveu (GN) or Gross-Neveu-Yukawa (GNY) transitions, which, as will be shown below, can be simulated via sign-problem-free QMC methods. 
For theoretical treatment, due to the absence of Fermi surfaces and the emergent Lorentz and conformal symmetry, some of the GNY QCPs are believed to be among the ``simplest'' fermionic QCPs, and thus hold the highest expectation for achieving agreement between controlled analytical calculations and numerical simulations. In recent years, such efforts have been attempted from both numerical and theoretical sides, e.g., the chiral-Ising GNY transitions have been studied via high-order expansions~\cite{Fourloopcritical,Ihrig2018} and quantum Monte Carlo simulations~\cite{Dynamicalgeneration,Quantumcriticalbehavior,TCLang2018,Schuler2019}.


In the numerical studies of GNY transitions and its quantum criticality, the most significant  challenge lies in the finite-size effect. A typical fermionic quantum Monte Carlo requires a complexity of $O(\beta N^3)$ where $N=L^{d}$ is the lattice size with $L$ being the linear span of the $d$ dimensional system and $\beta$ the inverse temperature. In addition to the typical challenge of critical slowing down, quantum critical points in the GNY models have two additional sources of finite-size effects. First, in a lattice model, Dirac fermions emerge in the close vicinity of the Dirac points. As far as critical phenomena are concerned, the part of the energy band near the Dirac point (with linear dispersion) contributes to the correct infrared (IR) scaling behavior, while the rest part of the energy band, which deviates from the Dirac linear dispersion, gives short-distance non-universal physics. In other words, the scaling behavior comes from only a portion of the Brillouin zone (less than $15\%$ in a typical lattice model as shown below), which effectively reduces the system size by $\sim 85\%$ in numerical simulations and thus significantly amplifies  finite-size effects. To enlarge the linear-dispersion region, special models are utilized, such as the SLAC fermion construction in the recent work of Ref.~\onlinecite{TCLang2018}. However, such special models often involve infinite-long-range hopping, which makes the system non-local.  Secondly, in a typically lattice model with bosons and fermions coupled together, these two types of particles in general have different velocities. This velocity difference explicitly breaks the Lorentz symmetry, and results in a very slow (logarithmic) RG flow of the velocities towards the fixed point~\cite{roy2016emergent}. Such a slow RG flow requires extremely large system sizes to reach the vicinity of the fixed point, making it very challenge to produce the correct exponents on finite size lattice.

In the study of fermionic quantum criticality, to overcome the critical slowing down, various efforts have been made to access larger system sizes, such as the self-learning Monte Carlo (SLMC)~\cite{SLMC2016,SLMC2017}, which provides more efficient updates, and its momentum space extension (elective momentum ultra-size Monte Carlo EQMC)~\cite{ZiHongLiuEMUS2018}. These improvements addresses the numerical aspect, and are implemented {\it a posteriori} to the model design. In this manuscript, we utilize a different approach to reduce the finite-size effect through {\it a priori} optimizing the model design to better utilize the emergent Lorentz symmetry at the chiral Ising GNY transition. In particular, we develop designer model Hamiltonians with two properties: (1) the Dirac linear-dispersion region occupies a larger portion of the Brillouin zone {\it without sacrificing locality} and (2) the bosons and fermions have the {\it same velocity} to avoid the slow RG flow~\cite{roy2016emergent}. These efforts minimize the finite-size effects mentioned above. Combined with advanced numeric scheme such as SLMC, we show explicitly that finite-size effects are efficiently suppressed in the obtained results, given rise to robust critical exponents. Furthermore, we find our designer Hamiltonian provide robust results in the behavior of free Dirac fermion conductivity at finite Matsubara frequency, such consistency gives one the confidence of obtaining the similar level analysis in the GNY QCP in the future works.  
 
The rest of the paper is organized as follows: in Sec.~\ref{sec:ii} the designer model of the $N_f=8$ chiral Ising GNY transition is introduced, as well as the techniques that we utilized to enlarge the linear dispersion region and to match the fermion and boson velocities. In Sec.~\ref{sec:iii}, the numerical results are presented, which include the crossing point analysis of the critical exponents (Sec.~\ref{sec:crossing}), the measurement of the two velocities at the GNY QCP (Sec.~\ref{sec:velocities}), and eventually the discussion of the behavior of the finite frequency conductivity of free Dirac fermion, as well as the conformal field theory prediction (Sec.~\ref{sec:conductivity}). Finally in Sec.~\ref{sec:iv}, the conclusions are drawn and few immediate future directions are outlined.

\begin{figure*}[htp!]
	\includegraphics[width =\textwidth]{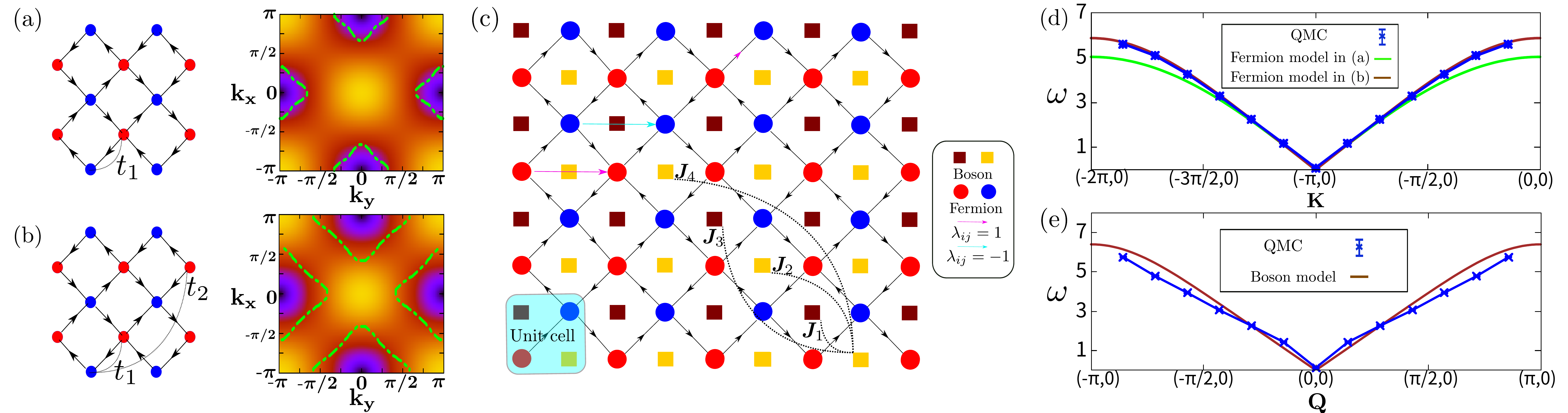}
	\caption{The fermion tight-binding models are shown in (a) and (b), with (a) only contains the nearest-neighbor hopping $t_1$ (Ref.~\cite{Dynamicalgeneration}) and (b) contain both the nearest-neighbor hopping $t_1$ and the third-nearest-neighbor hopping between different sublattices $t_2$ (this work). The green dashed lines in the right panels mark the linear dispersion region, out of which the band structure starts to deviate from the Dirac linear dispersion. The $t_2$ hopping in Fig.~(b) greatly increases the area of the linear dispersion region. 
The black arrows in the left panel represent the phase of the complex hopping strength $\theta=\frac{\pi}{4}$, and we set $t_1=2$ and $t_{2}=\frac{t_{1}}{27}$. The value of $t_2$ is obtained from the optimization process of trying to enlarge the linear dispersion area in BZ. The lattice model is depicted in (c), where the coupling constants for bosons are $J_{2}=-\frac{J_{1}}{8}, J_{3}=\frac{J_{1}}{63}, J_{4}=-\frac{J_{1}}{896}$. The values of boson couplings are also obtained from the parameter optimization process. The bosons and fermions are coupled together via the $\lambda_{i,j}=\pm 1$  term shown in Eq.~\eqref{eq:eq2}, where two next-nearest-neighbor fermions are coupled together with the boson between the two fermion sites. The sign
of the coupling constant $\lambda_{i,j}$ alternates from one plaque to another, as marked by in (c). Overall, one unit cell of this model contains two fermion sites and two boson sites. (d) and (e) show dispersions of fermions and bosons respectively (see Appendix~\ref{app:appC} for detail explanation of how to obtain these dispersions). Here, we presented both the bare dispersions (ignoring interactions) and the renormalized ones at the chiral Ising transition via QMC simulations (with $L=14$). In panel (d), it is easy to notice that model shown in panel (b) greatly increases the linear-dispersion region.} 
	\label{fig:fig1}
\end{figure*}

\section{Designer model}
\label{sec:ii}
The designer model is realized by coupling Dirac fermions with dynamical bosonic field~\cite{Dynamicalgeneration,Quantumcriticalbehavior,UniversalQuantumCriticality}. 
The model is defined on a square lattice shown in Fig.~\ref{fig:fig1}, with two bosonic sites and two fermionic sites per unit cell. The bosons acquire an Ising symmetry and the 
Lagrangian of the system is given as 
\begin{equation}
\label{eq:eq1}
L=L_{\text{Boson}}+L_{\text{Fermion}}+L_{\text{Coupling}}
\end{equation}
with
\begin{eqnarray}
	\label{eq:eq2}
	\begin{split}
	&L_{\text{Boson}}=\sum_{p}\left[\frac{1}{4}\big(\frac{\partial\phi_p}{\partial \tau}\big)^2+m\phi^2_p+\phi^4_p \right]+\sum_{(p,q)}J_{pq}(\phi_p-\phi_q)^2,\\
	&L_{\text{Fermion}}=\sum_{( i, j ),\sigma}\psi^{\dagger}_{i,\sigma}[(i \partial_{\tau}-\mu)\delta_{ij}-t_{ij}\text{e}^{i\sigma\theta_{ij}}]\psi_{j,\sigma} + h.c., \\	
	&L_{\text{Coupling}}=\sum_{\langle \langle i,j \rangle \rangle,\sigma}\lambda_{ij}\phi_{p}\psi^{\dagger}_{i,\sigma}\psi_{j,\sigma} + h.c.,
\end{split}
\end{eqnarray}
where $\psi_{i,\sigma} \ (\psi^\dagger_{i,\sigma})$ is the fermionic annihilation (creation) operator at the fermionic site $i$ with spin $\sigma=\uparrow$ or $\downarrow$, and $\phi_{p}$ represents the scalar bosonic field at the bosonic site $p$. $\mu$ is chemical potential and we set $\mu=0$ to ensure the half-filling of fermion. 

Fig.~\ref{fig:fig1} (a) and (b) show the fermionic part of the model.
The fermionic hopping strength carries a complex phase $\theta_{ij}$. For the nearest-neighbor hopping between different sublattices $t_1$ and the third-nearest-neighbor hopping between different sublattices $t_2$, the phase factor is set to $\theta=\pi/4$, which introduces a $\pi$-magnetic-flux for each plaquette of the fermion lattice.
This tight-binding model gives rise to two Dirac cones at $(\pi,0)$ and $(0,\pi)$ in the Brillouin zone (BZ). The conventional $\pi$-flux model with only $t_1$ is depicted in Fig.~\ref{fig:fig1} (a), which is the tight-binding model employed in previous studies~\cite{Dynamicalgeneration,Quantumcriticalbehavior}. The model utilized in this work is shown in Fig.~\ref{fig:fig1} (b), where one additional hopping term $t_2$ is introduced with $t_2=t_1/27$. As shown in the right panels of Fig.~\ref{fig:fig1} (a) and (b), this additional hopping term greatly increases the area of the linear dispersion region, denoted by the green contour lines, from 13.7$\%$ in (a) to 39.8$\%$ in (b) of the entire BZ. Here, this green contour marks momentum points, at which the energy band deviates from the ideal linear dispersion by 5$\%$. It serves as the ultraviolet cutoff ($\Lambda$) in the renormalization group (RG) analysis of this Dirac system. 

For a critical system, the RG flow is governed by the dimensionless quantity $L\Lambda$, where $L$ is the system size. A larger $\Lambda$ effectively increases the system size, and thus pushes the system closer to the thermodynamic limit. Similar ideas of increasing the linear dispersion area  has been used in another recent QMC study~\cite{TCLang2018}, where the linear dispersion region is expanded to the entire BZ via introducing infinite-long-range hopping terms, known as the SLAC fermion model. These long-range hopping terms render the model non-local. 
In contrast, in the model that we adopted here, although the linear-dispersion region doesn't cover the entire BZ, the Hamiltonian is local with only short-range couplings.

Fig.~\ref{fig:fig1} (c) depicts the full model with circles representing the fermionic sites and squares the bosonic ones. Similar to the fermions, the bosons here also have extended, but local, interactions, denoted as $J_1$, $J_2$, $J_3$ and $J_4$ from the first-neighbor to the fourth-neighbor couplings respectively. The values of these coupling strengths are shown in the caption of Fig.~\ref{fig:fig1}. At the QCP, the critical boson mode exhibits a linear dispersion at small wavevector.  These values of the $J$'s maximize the area of this linear-dispersion region, which helps to suppress finite-size effects same as what we did for the fermions above. With this parameter choice, the bare boson velocity at the critical point of $L_{\text{Boson}}$ is $v_{b}=2\sqrt{5J_{1}/8}$, and the Fermi velocity at the Dirac point is $v_f = \frac{8}{9}\sqrt{2}t_1$. Here, we set $t_1=2$ and choose the value of $J_1$ such that bosons and fermions share identical velocity $v_b=v_f$ (See Appendix~\ref{app:appA} for details). This identical velocity results in an emergent Lorentz symmetry, where $v_b=v_f$ serves as the speed for light. For a Lorentz-invariant system, it is known that the speed of light would not flow under RG, even in the presence of strong interactions~\cite{Weinberg2005}. This {\it a priori} setup avoids the slow logarithmic RG flow of the velocity mentioned above~\cite{roy2016emergent}.

The couplings between bosons and fermions are denoted as $\lambda_{ij}$, which couples to a pair of next-nearest-neighbor fermion sites with the boson site between them. The coupling constant $\lambda_{ij}$ takes the value of $\pm 1$, where the two bosonic sublattices take opposite signs. In the ordered phase (i.e. the boson field has a nonzero expectation value), this alternating sign gives rise to the required Berry phase to the fermions, and thus gaps out the Dirac points, transforming the Dirac semi-metal into a dynamically-generated quantum-spin-Hall insulator. More details about such a topological phase transition and numerical realization can be found in Ref.~\cite{Dynamicalgeneration}. 

Fig.~\ref{fig:fig1} (d) and (e) depict the dispersions near the fermionic Dirac point at $(\pi,0)$ and bosonic dispersion at $(0,0)$ at the QCP. The dispersions are obtained from the fitting of the imaginary time fermionic and bosonic Green's functions, and are explained in Appendix~\ref{app:appC}. The green and brown lines in Fig.~\ref{fig:fig1} (d) are the bare fermion dispersions for the models shown in Fig.~\ref{fig:fig1}(a) and (b), respectively. One can see that the linear-dispersion region for the brown line is indeed larger than that of the green line. The blue stars are the QMC results obtained at the chiral Ising critical point. Fig.~\ref{fig:fig1} (e) presents the bosonic dispersion at the bare Ising critical point (the brown line) and the chiral Ising critical point (the blue stars). 
It is worthwhile to highlight that for both fermions and bosons, their velocities receive little renormalization. This absence of velocity renormalization is because we set the bosons and fermions to have the same bare velocity as mentioned above (we think the deviation of the boson dispersion at the QCP from the brown line still comes from the inevitable finite size effect of $L=14$ lattice). 
More details about the RG flow of the velocities is provided below in the next section.

To solve the model shown in Eq.~\eqref{eq:eq1}, in particular to obtain its ground state phase diagram and dynamic properties, we employ the  projective QMC method~\cite{Assaad2008,Meng2010,Dynamicalgeneration}  with SLMC update schemes. Detailed information of the numeric implementation, in particular on how to obtain accurate effective model in the SLMC within the projective QMC framework, is given in Appendix~\ref{app:appB}.

\begin{figure}[htp!]
		\includegraphics[width=\columnwidth]{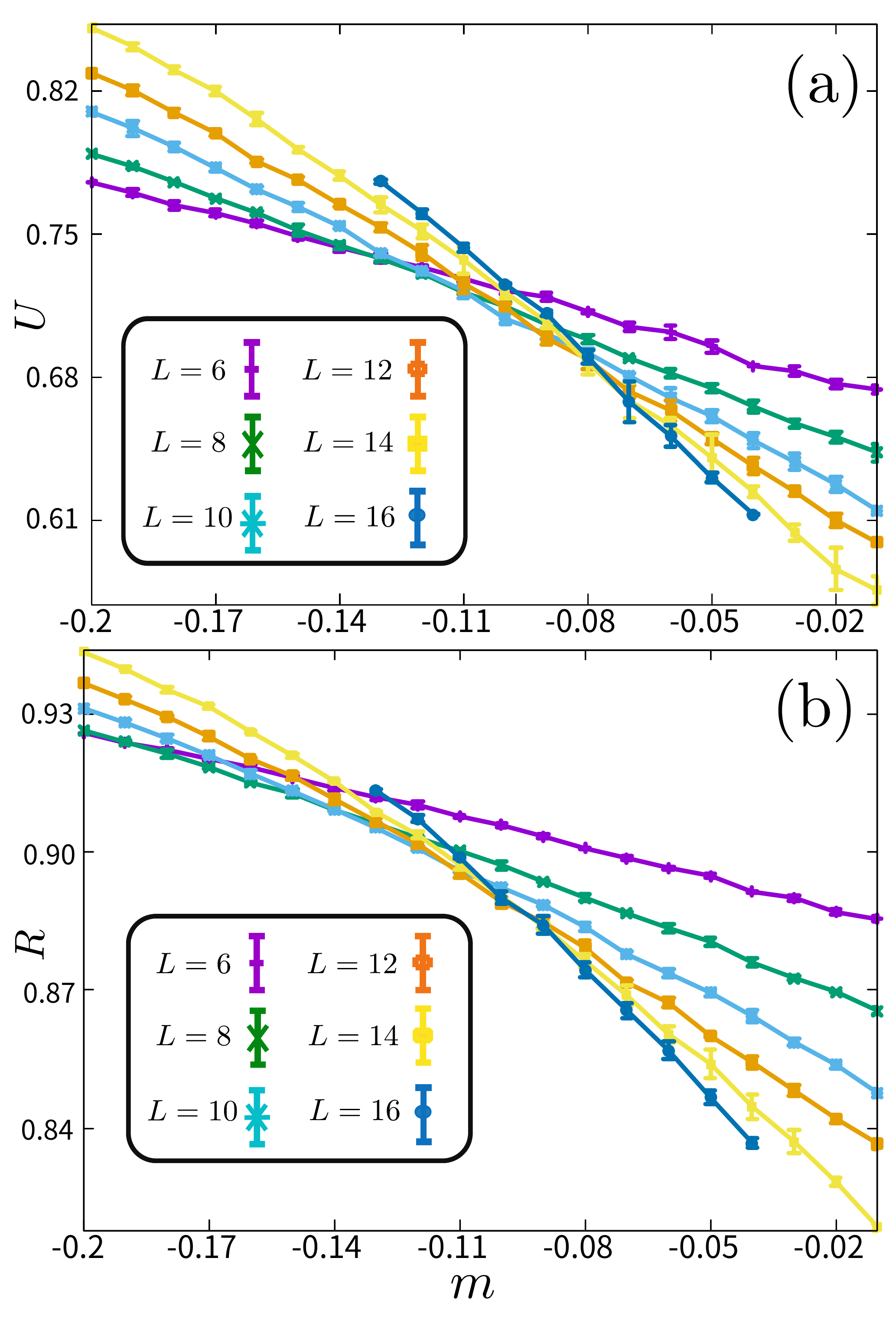}
		\caption{(a) The Binder ratio $U(L,m)$ for different system sizes $L$ across the transition of $m$. Curves for two consecutive system sizes $L$ and $L+2$ cross at the finite size transition point $m_c(L)$ . (b) The correlation ratio $R(L,m)$ for different system sizes $L$ across the transition of $m$. Curves for two consecutive sizes $L$ and $L+2$ cross at the finite size transition point $m_c(L)$. The finite-size analysis for these crossing points are shown in Fig.~\ref{fig:fig3}.}
		\label{fig:fig2}
	\end{figure}

\section{Results}
\label{sec:iii}
In the simulation, we vary the value of the mass term $m$ in $L_{\text{Boson}}$ in Eq.~\eqref{eq:eq1}.
Without fermions, tuning this mass term triggers a $(2+1)$d Ising transition for the $\phi$ field. In the presence of the fermions, this transition induces a fundamental change in the fermion band structure. In the disordered (ordered) phase, $\langle \phi \rangle =0$ ($\langle \phi \rangle \ne 0$), the Dirac fermions is massless (massive), i.e., the long-range order in bosons generates finite mass for the Dirac fermions. In addition, it can be shown that this Dirac mass carry a nontrivial topological index, and results in a dynamically generated quantum spin Hall insulator~\cite{Dynamicalgeneration}. Furthermore, the interplay between bosons and fermions also changes the universality of the quantum critical point, from Ising to  the GNY chiral Ising, which is believed to be among the simplest fermionic QCPs where possibly analytical and numerical approaches might be able to give consistent critical exponents to build a concrete CFT description~\cite{Fourloopcritical,Iliesiu2018,Rong2018,Schuler2019}. For the model described above, the Dirac fermions acquire $N_f=8$ flavors, and thus the QCP belongs to the $N_f=8$ GNY chiral Ising universality class. Below, we reveal step by step how the critical properties of this transition is obtained in our designer QMC simulations.

\begin{figure}[htp!]
	\includegraphics[width =\columnwidth]{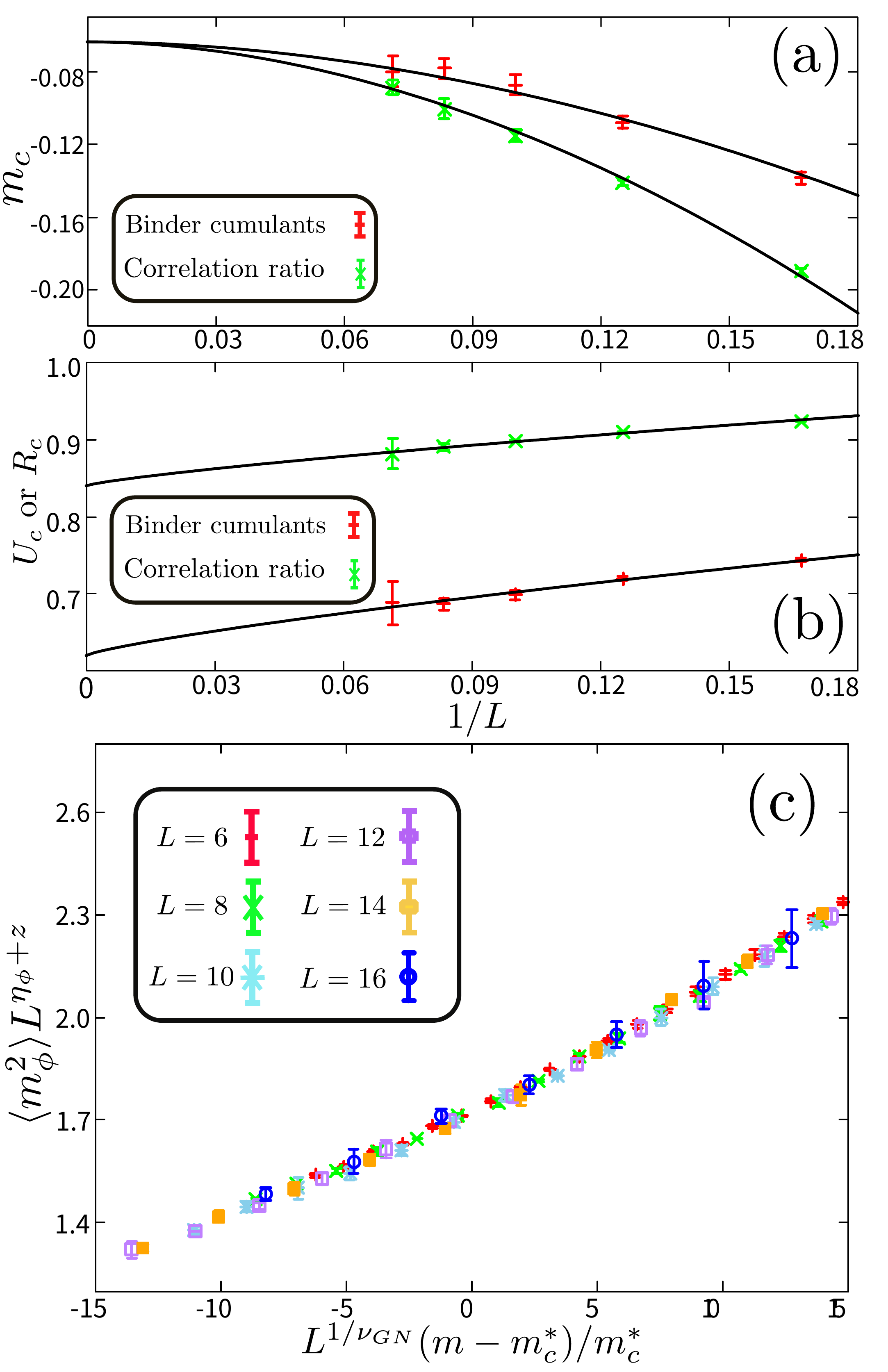}
	\caption{(a) Crossing point analysis for the Binder and correlation ratios obtained in Fig.~\ref{fig:fig2}. The two $m_{c}(L)$ curves extrapolate to the same critical point $m^{*}_{c}=- 0.062(3)$ at the thermodynamic limit (note that we set the two curves to meet at a single value the $m^{*}_c$ in the fitting) and the two powers in the extrapolation according to Eq.~\eqref{eq:eq6} consistently reveal $1/\nu_{\text{GN}}+\omega=1.8(1)$ , with different non-universal coefficients $a$. (b) Crossing point analysis of the Binder ratio $U_c(L)$ and the correlation ratio $R_c(L)$ according to Eqs.~\eqref{eq:eq7} and \eqref{eq:eq8}. The obtained correction exponent $\omega = 0.8(1)$ . Combining the exponents from (a) and (b), we find $1/\nu_{\text{GN}}=1.0(1)$. (c) Data collapsing for the order parameter measured at different $m$ and system sizes $L$. Here, we utilized the the exponent $\nu_{\text{GN}}$ and the critical point $m^{*}_{c}$ obtained above and the system sizes are $L=6, 8, 10, 12, 14, 16$. A good collapse is achieved with dynamic critical exponent $z=1$ and the anomalous dimension $\eta_{\phi}=0.59(2)$.}
	\label{fig:fig3}
\end{figure}

\subsection{Crossing-point analysis}
\label{sec:crossing}
First we determine the precise location of the quantum critical point via the Binder ratio ($U$)~\cite{Binder1981} and correlation ratio ($R$)~\cite{Pujari2016} of the bosonic order parameter 
\begin{eqnarray}
\label{eq:eq3}
m^2_{\phi}  = \frac{1}{N}\sum_{\langle p,q \rangle}\phi_{p}\phi_{q}
\end{eqnarray}
where $N$ is the number of boson lattice sites, and the two dimensionless ratios are given as 
\begin{align}
\label{eq:eq4}
U&=\frac{3}{2}\left(1-\frac{1}{3}\frac{\langle m_{\phi}^4 \rangle }{\langle m_{\phi}^2 \rangle^2}\right)\\
R&=1-\frac{S(\mathbf{Q}+\delta_{\mathbf{q}})}{S(\mathbf{Q})}
\end{align}
here $S(\mathbf{q})=\frac{1}{N}\sum_{\langle p,q \rangle} e^{-i\mathbf{q}\cdot(\mathbf{r}_p-\mathbf{r}_q)} \phi_{p}\phi_{q}$ is the structure factor of the bosonic correlation function and  
$\mathbf{r}_p$ and $\mathbf{r}_q$ stand for the position of the bosonic field on the lattice in Fig.~\ref{fig:fig1} (c). The ordering wavevector in our case is $\mathbf{Q}=(0,0)$, and $\delta_{\mathbf{q}}=(0,\pi/L)$ or $(\pi/L,0)$ is the smallest momentum away from $\mathbf{Q}$ on a finite lattice with linear span $L$.

The results are shown in Fig.~\ref{fig:fig2} (a) and (b). The finite-size transition point $m_c(L)$ is the crossing points of the two consecutive system sizes, and it is clear from the two panels of Fig.~\ref{fig:fig2} that the position of the $m_{c}(L)$-s are gradually drifting as $L$ increases. According to the finite size scaling analysis~\cite{Shao2016,YQQin2017}, the drift of both the critical point $m_c(L)$ and the corresponding $R_{c}(L)$ and $U_{c}(L)$ obey the following scaling behavior at near the critical point,
\begin{align}
\label{eq:eq6}
m_{c}(L)&=m^{*}_{c}+aL^{-(1/\nu_{\text{GN}} + \omega)}, \\	
\label{eq:eq7}
U_{c}(L)&=U^{*}_{c}+bL^{-1/\omega},\\
\label{eq:eq8}
R_{c}(L)&=R^{*}_{c}+cL^{-1/\omega},
\end{align}
where $\nu_{\text{GN}}$ and $\omega$ are the correlation length exponent and the correction expoent of the chiral Ising GNY transition. $m^{*}_{c}$,  $U^{*}_{c}$ and $R^{*}_{c}$ are critical point, correlation ratio and Binder ratio in the thermodynamic limit and $a$, $b$ and $c$ are non-universal fitting coefficients.

\begingroup
\setlength{\tabcolsep}{6pt} 
\renewcommand{\arraystretch}{1.5} 
\begin{table*}[htp!]
	\centering
	\begin{tabular}{c c c c c}
		\hline\hline
		\ & $1/\nu_{\text{GN}}$ & $\eta_{\phi}$ & $\eta_{\psi}$ & $\omega$ \\
		\hline
		This work & 1.0(1) & 0.59(2) & 0.05(2) & 0.8(1) \\
		previous QMC~\cite{Dynamicalgeneration}& 1.2(1) & 0.65(3) &  &     \\
		previous QMC~\cite{Quantumcriticalbehavior}&1.20(1) & 0.62(1) & 0.38(1) & \\
		perturbative RG 4-$\epsilon$ loop, Pad\'e ~\cite{Fourloopcritical}& 0.931 & 0.7079 & 0.0539 & 0.794 \\ 
		perturbative RG 4-loop, interpolation betwen 2+$\epsilon$ and 4-$\epsilon$~\cite{Ihrig2018}&1.004 & 0.735 & 0.042& \\
		bootstrap estimation~\cite{Iliesiu2018}& 0.88 & 0.742 & 0.044 & \\
		\hline\hline
	\end{tabular}
	\caption{Comparison of obtained critical exponents for the $N_f=8$ chiral Ising GNY universality class. The results are obtained from QMC simulations, 4-loop perturbative RG and the estimate of conformal bootstrap bound. $\nu_{\text GN}$ is the correlation length exponent, $\eta_{\phi}$ is the anomalous dimension of the bosonic field, $\eta_{\psi}$ is the anomalous dimension of the fermionic field and $\omega$ is the correction exponent in the bosonic sector.}
	\label{tab:tab1}
\end{table*}
\endgroup

As shown in Fig.~\ref{fig:fig3}(a), the crossing points of different pairs of system size $(L,L+2)$ drift towards $m^{*}_{c}$ as a power-law function as shown in Eq.~\eqref{eq:eq6}. It is important to emphasize here that for both the Binder ratio $U$ and the correlation ratio $R$, the scaling analysis generates the same critical point $m^{*}_{c} = -0.062(3)$ and the same critical exponents $1/\nu_{\text{GN}}+\omega = 1.8(1)$. This consistency indicates that finite-size effects are under good control in the current simulation, even though our system sizes (upto $L=16$) are just moderately large.

\begin{figure}[htp!]
	\includegraphics[width =\columnwidth]{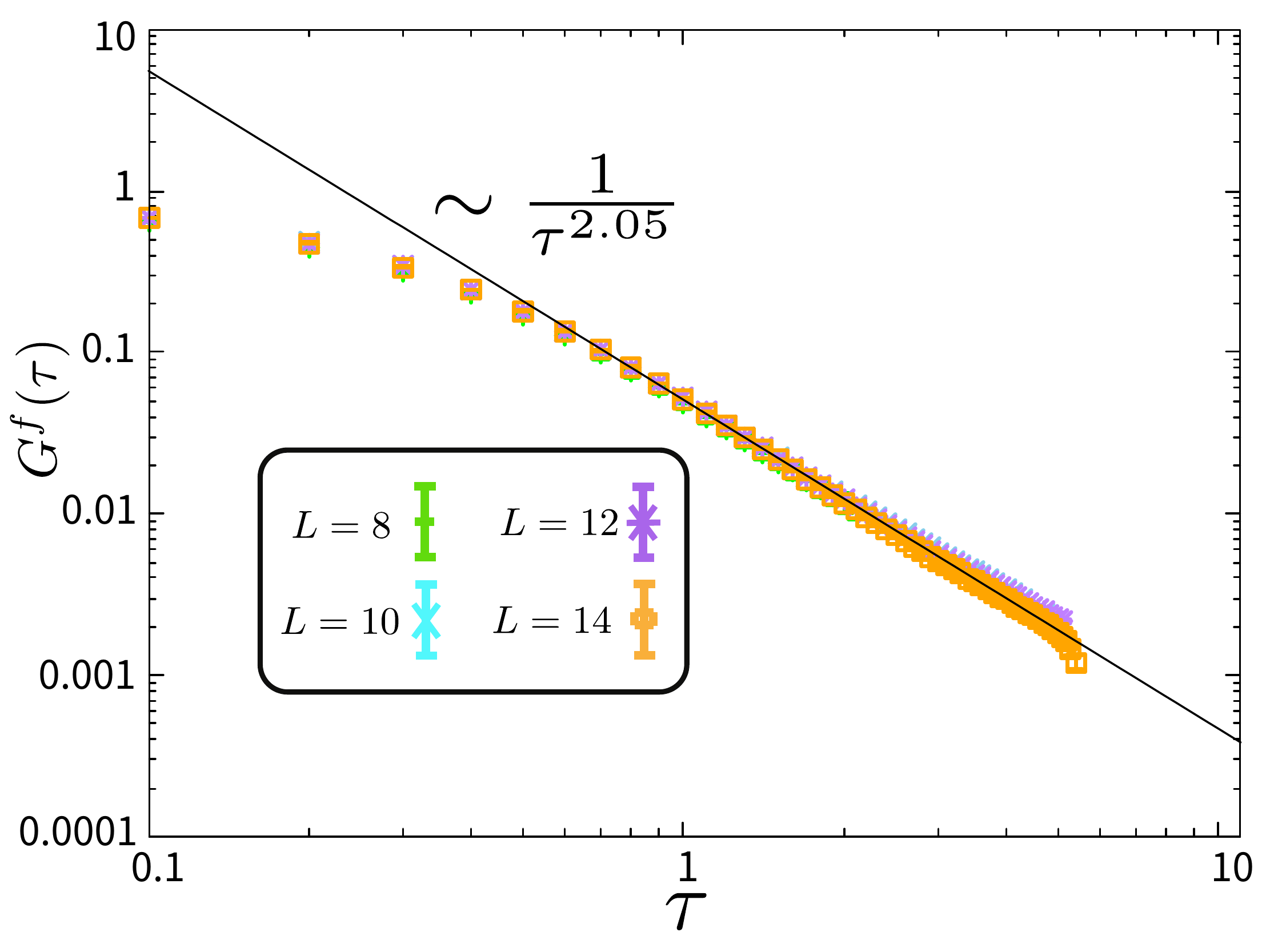}
	\caption{The imaginary time decay of the fermion Greens's function at the chiral Ising QCP. From the power-law decay of $G^{f}(\tau) \sim 1/\tau^{2+\eta_{\psi}}$ we obtain the anomalous dimension of the Dirac fermions $\eta_{\psi}=0.05(2)$.}
	\label{fig:fig4}
\end{figure}

In Fig.~\ref{fig:fig3}(b), we fit the two curves according to Eqs.~\eqref{eq:eq7} and \eqref{eq:eq8}, to trace the universal thermodynamic values of the Binder and correlation ratios $U^{*}_{c}$ and $R^{*}_{c}$, as well as the correction exponent $\omega$. It turns out that both the coefficient $b$ and $c$ are positive numbers (in general, their signs depend on the model and could be opposite in certain systems as shown in Ref.~\cite{YQQin2017}), such that both curves decreases as $L$ increases. Interestingly, the two independent fits (for $U$ and $R$) yields the same same exponent value $\omega=0.8(1)$. With the value of $\omega$ obtained, we can now determine $1/\nu_{\text{GN}}=1.0(1)$ from the combined exponents in Fig.~\ref{fig:fig3} (a).

With $m^{*}_c$, $\nu_{\text GN}$, we can directly collapse the bosonic order parameter with the universal scaling form $\langle m_{\phi}^2 \rangle L^{z+\eta}=f(L^{1/\nu_{\text{GN}}}(m-m^{*}_{c})/m^{*}_{c})$. The results are shown in Fig.~\ref{fig:fig3} (c), with $z=1$ of the $(2+1)$D chiral Ising universality, the data of $L=6,8,\cdots, 16$ give a very good collapse and from which we read the bosonic anomalous dimension $\eta_\phi=0.59(2)$ at the chiral Ising GNY QCP.

The fermionic anomalous dimension $\eta_\psi$, however, cannot be obtained from the above bosonic crossing analysis but have to be extracted from the fermion Green's function at the QCP. For a QCP with the dynamic critical exponent $z=1$, there are two ways to obtain this anomalous dimension, by fitting either the time dependence or the spatial dependence of the Green's function, and they should produce the same value. But since in the lattice simulation the spatial and time axes are actually anisotropic, one has more distance along the time-axis than that of the spatial-axis. Moreover, since the Lorentz symmetry in a lattice model is an emergent symmetry instead of an explicit one, time and space are not fully identical in our model. In our previous work~\cite{Dynamicalgeneration}, with only nearest-neighbor fermion hopping, we could not obtain power-law decay in real-space of the fermion Green's function at the finite size QCP. In this work, although the results have been greatly improved, we still see that for the fermionic anomalous dimension, the time-dependence data has more data points and exhibiting better power-law behavior than the space-dependence data, and we thus perform a more accurate fitting with the time-dependence data of the fermion Green's function. 

In Fig.~\ref{fig:fig4}, we present the imaginary time decay of the fermion Green's function $G^f(\tau)=\frac{1}{N}\sum_{i=1}^{N}\langle \psi^{\dagger}_{i,\sigma}(\tau)\psi_{i,\sigma}(0)\rangle$ at the QCP $m=m^{*}_c$ and fit the data with its scaling form
\begin{equation}
G^f(\tau) \sim 1/\tau^{2+\eta_\psi},
\label{eq:fermioneta}
\end{equation}
where the power 2 is the bare-scaling dimension for free Dirac fermions in $(2+1)$D, and $\eta_\psi$ accounts for the anomalous part of the exponent at the nontrivial GNY fixed point. In Fig.~\ref{fig:fig4}, we plot $G^f(\tau)$ at the QCP ($m=m^{*}_c$) for several system sizes in the log-log scale. It is clear that at large imaginary time, the power-law decay of the fermion Green's function manifests with an anomalous dimension $\eta_\psi=0.05(2)$.

This completes our analysis of the critical exponents of the $N_f=8$ chiral Ising GNY transition. In Tab.~\ref{tab:tab1}, we list the obtained critical exponents and compare them with previous QMC works~\cite{Quantumcriticalbehavior,Dynamicalgeneration} and various high-order perturbative RG results~\cite{Fourloopcritical,Ihrig2018} as well as conformal bootstrap estimations~\cite{Iliesiu2018} (we note here the conformal bootstrap only gives estimation of the exponents based on the position of the kinks from bootstrap calculations~\cite{Iliesiu2018,Rong2018}, the exact value of the exponents within conformal bootstrap analysis for the fermion GNY transitions are yet to be obtained).  

Two points are in order concerning the obtained exponents in this work in comparison with previous numerical studies. First, in the previous works, in particular the QMC works, finite-size effects have not been reduced as much as in the current designer Hamiltonian, i.e., the fermion and boson velocities may not be the same at the bare level and the RG flow might not be able to synchronize them limited by the finite system size, whereas in the present case, the $v_f = v_b$ \textit{a priori} and as will be shown in the next section, we find that this setup avoided the slow RG flow of velocity, such that the finite-size simulations are closer to the nontrivial fixed point of the QCP. Second, the $1/\nu_{\text{GN}}$, $\eta_{\phi}$ in Tab.~\ref{tab:tab1} are consistent with previous QMC results, and because finite-size effects have been efficiently supressed in the present study, although our system sizes are only moderately large ($L=16$), we can perform the crossing-point scaling to independently obtain $1/\nu_{\text{GN}}$, $\eta_{\phi}$ and the correction exponent $\omega$, where the later cannot be accessed in the previous QMC simulations. Our fermionic anomalous dimension $\eta_{\psi}$ is different from the the previous QMC result~\cite{Quantumcriticalbehavior}, and is significantly closer to the theory predictions from perturbative RG and conformal bootstrap. In our previous QMC work~\cite{Dynamicalgeneration}, the $\eta_{\psi}$ couldn't be obtained through the fitting formula shown in Eq.~\eqref{eq:fermioneta}, precisely because the strong finite-size effect in the simulation renders the slow (logarithmic) RG flow and consequently the Green's function data at finite sizes cannot produce a robust value of $\eta_{\psi}$. Therefore we are more confident with $\eta_\psi$ obtained in this work. In fact, the exponents reported here, $\{1/\nu_{\text{GN}}, \ \eta_{\phi}, \ \eta_{\psi}, \ \omega\}$ offer the most complete set of numerically measured critical exponents till now for the $N_f=8$ chiral Ising GNY transition. Thanks to the careful model design, these results provide the much needed benchmark for future numerical and analytical investigations.

\subsection{Velocities at transition point}
\label{sec:velocities}
As have been mentioned above, our designer Hamiltonian enjoys the advantage of maintaining the Lorentz symmetry even at the bare level, i.e., the linear dispersion region of both fermion and boson are enlarged in the BZ and their velocities are made identical by choosing suitable values of $t_1$ and $J_1$. With such optimization, finite-size effects at the chiral Ising GNY QCP are reduced and thus the exponents converges to stable values even with just moderately large system sizes up to $L=16$.
To further verify whether the fermion and boson velocities indeed remain identical and maintain their bare values, here we directly measure these two velocity at the chiral Ising GNY QCP.

Here, we measure the velocities via the dynamical Green's functions (the dynamical fermionic and bosonic propagators) $G^{f}(\mathbf{k},\tau)=\langle c^{\dagger}(\mathbf{k},\tau)c(\mathbf{k},0)\rangle$ and $G^{b}(\mathbf{q},\tau)=\langle \phi(\mathbf{q},\tau)\phi(-\mathbf{q},0)\rangle$. In the projector QMC, these two Green's functions are readily obtained from the imaginary time correlation functions~\cite{Assaad2008,Meng2010}. Since both fermion and boson have two sublattices, we measure the trace of the $2\times 2$ matrices at each momentum point, and from fitting the imaginary time decay of $G(\mathbf{k},\tau) \propto e^{-\Delta(\mathbf{k})\tau}$, the many-body excitation gaps are obtained at each momenta $\mathbf{k}$ and $\mathbf{q}$. This relation between the excitation gap and the momentum marks the ``mass shell'', in analogy to the relativistic quantum field theory.
Then we fit this relation near the Dirac point $\mathbf{K}=(\pi,0)$ for fermions and near the $\Gamma$ point $\mathbf{Q}=(0,0)$ for bosons as $\Delta_{f} \sim v_{f}(\mathbf{k}-\mathbf{K})$ and $\Delta_{b} \sim v_{b}(\mathbf{k}-\mathbf{Q})$ respectively. This is how the velocities at the strong-coupling fixed point are obtained(see Appendix \ref{app:appC} for detail calculation). 

The results are shown in Figs.~\ref{fig:fig1} (d) and (e). With system size up to $L=14$, we find that $v_f$ is very close to its bare value with very good linear behavior and $v_b$ is also close to its bare value with slightly larger error-bars and deviations. This is due to the fact that our boson is a continuous field and the Monte Carlo moves, even with non-local SLMC update scheme, still suffers from the slow Monte Carlo dynamics and rendering slightly worse data quality. But, nevertheless, As shown in Fig.~\ref{fig:fig1} (d) and (e), $v_f$ does't flow and remains a constant, in agreement with theory expectation.  The boson dispersion suffers more from finite-size effect and slow QMC updates as shown in Fig.~\ref{fig:fig1} (e).  Within the uncertainty of finite-size effects, it is believed to consist with theory expectation. However, to verify the theory prediction at the quantitative level, larger systems size are needed to overcome the finite-size effect and achieve the thermodynamic limit. This result confirms the absence of the slow velocity RG flow, which helps control the finite-size effect in our simulations. 

\begin{figure}[htp!]
\includegraphics[width =\columnwidth]{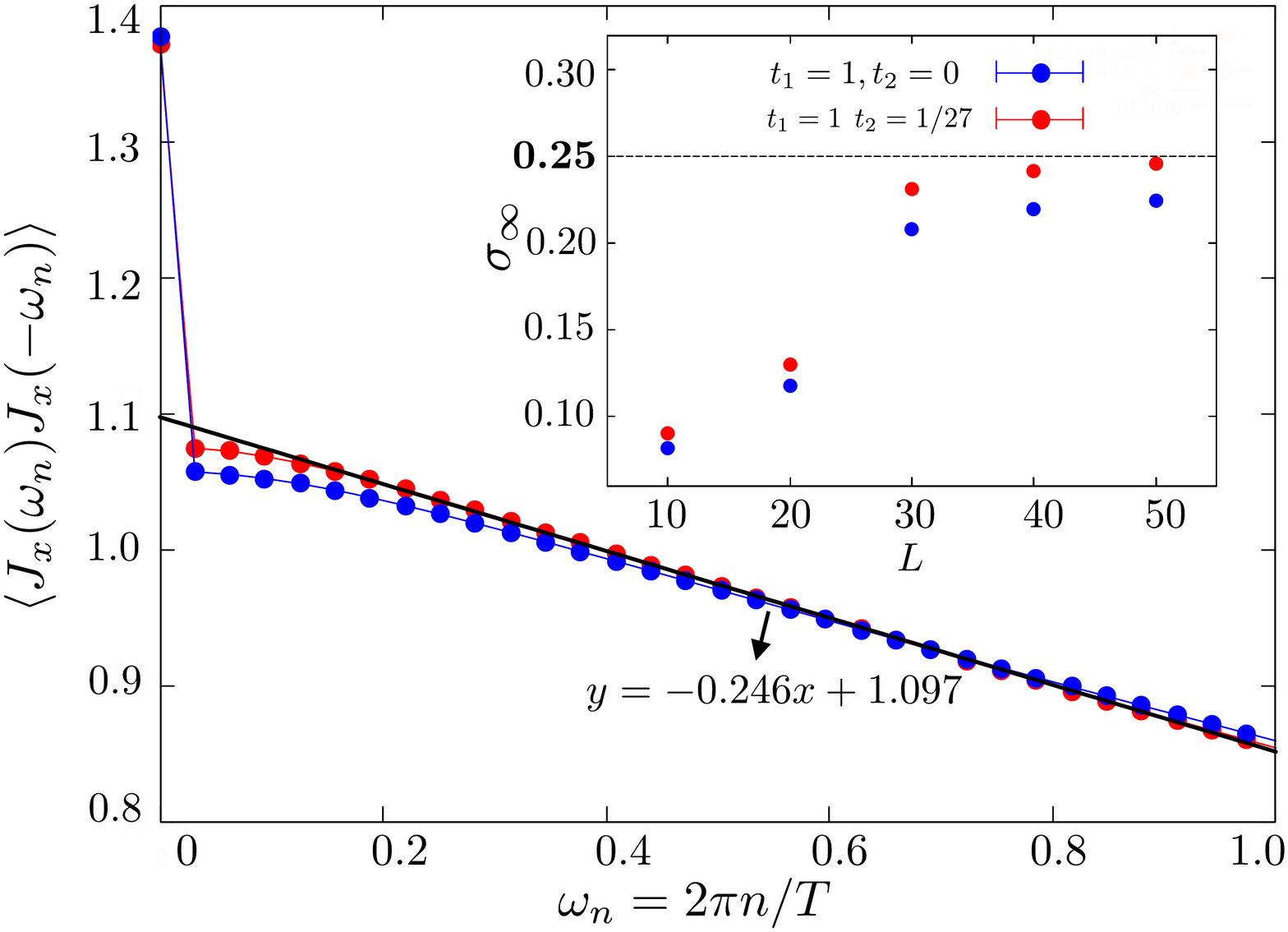}
\caption{The frequency dependence of the current-current correlation with $L=50$ and $T=1/200$ for free Dirac fermions, where the red (blue) data points are obtained from the designer model with $t_2=1/27$ (the conventional model with $t_2=0$).
The black solid line is the fitting to the conformal theory prediction [Eq.~\eqref{eq:eq11}], which gives rise to $\sigma_{\infty} = \frac{1}{4}$ at the limit of $\omega_n \gg T$. 
The inset shows the fitted $\sigma_{\infty}$ as a function of the system size $L$. As $L$ increases,  $\sigma_{\infty}$ from the designer model converges to the exact value (the dash line) much faster than that of the conventional model.}
\label{fig:fig5}
\end{figure}

\subsection{Dirac fermion conductivity}
\label{sec:conductivity}
With the basic scaling properties of the $N_f=8$ chiral Ising GNY QCP determined, we can now pursue more sophisticated properties of this fermionic QCP, e.g. the conductivity. In condensed matter physics, conductivity is one of the most commonly investigated physical quantities for both the theoretical calculation and experimental measurement. However, systematic studies of the scaling behavior of the conductivity at the close vicinity of a fermionic quantum critical point remains a challenging topic. From the current-current correlation which can be computed in QMC simulation, it is possible to extract the information of conductivity. In particular, it is proposed that the conductivity at the GNY QCP has strong connections with the conformal field theory~\cite{Katz2014,Krempa2014}, and subsequently demonstrated in bosonic $(2+1)$D quantum rotor models at the Wilson-Fisher fixed point~\cite{Krempa2014,Katz2014,Chen2014,Lucas2017} that the quantum critical conductivity at finite Matsubara frequency indeed obeys the scaling law predicted by the conformal field theory. However, similar results in fermionic QCPs are still absent.

From previous studies of the $(2+1)$D O(2) transition, it is known that very large $L$ and $\beta$ are needed to observe the correct scaling behavior~\cite{Krempa2014,Katz2014,Chen2014,Lucas2017}
\begin{eqnarray}
\frac{\sigma(i\omega_n)}{\sigma_{Q}}&=&-\frac{1}{\omega_n}\langle J_{x}(\omega_n)J_{x}(-\omega_n)\rangle + \cdots  \nonumber\\
 &=& \sigma_{\infty}+ b_{1}\big(\frac{T}{\omega_n}\big)^{3-1/\nu} + b_{2}\big(\frac{T}{\omega_n}\big)^{3} + \cdots,  \label{eq:eq10}
\\
 && \text{at} \;\; \omega_n \gg T \nonumber
\end{eqnarray}
where $\sigma(i\omega_n)$ is the conductivity at the Matsubara frequency $\omega_n=2\pi nT$, and $\sigma_{Q}=\frac{(e^*)^2}{\hbar}$ is the quantum unit of conductance with $e^*$ the effective charge in the model such that $\sigma/\sigma_{Q}$ becomes dimensionless. To simplify the discussion, we will set $\sigma_{Q}=1$ in the following discussion.  The current operator $J_{x}(\omega_{n})$ in Eq.~\eqref{eq:eq10} is the Fourier transformation of $J_x(\mathbf{r}_i,\tau)=-it \langle (e^{-i\sigma \theta}c^{\dagger}(\mathbf{r}_i,\tau)c(\mathbf{r}_{i+\hat{x}},\tau)-e^{i\sigma \theta}c^{\dagger}(\mathbf{r}_{i+\hat{x}},\tau)c(\mathbf{r}_{i},\tau))\rangle $ where $\hat{x}$ denotes the current along the  $x$-direction of the lattice. We note that here the $\hat{x}$ and $\hat{y}$ directions are equivalent due to the four-fold rotational symmetry, and thus can be symmetrized in the simulation to increase the data quality.  $\sigma_{\infty}$ is the limiting value of conductivity at $T \to 0$, $b_{1}$ and $b_{2}$ are coefficient related with the operator product expansion of the current operators in terms of other operators of the CFT and can in principle be estimated in the large $N$ limit. Coefficients at bosonic $O(N)$ Wilson-Fisher CFT have been estimated at large $N$~\cite{Katz2014} and also recent conformal bootstrap results for $N=2$~\cite{Chester2019}, but those at fermionic QCPs such as the current one have not yet been calculated to the best of our knowledge. 

Because the system size in our QMC simulations is only moderately large, $L=16$, we will restrain ourselves from the computation of $\sigma(i\omega_n)$ at the GNY chiral Ising QCP and leave it for a separated project. Here we will only demonstrate the computation for free Dirac fermions in the absence of interactions, i.e., bare fermions at the tree level, such that large $L$ and $\beta$ can be accessed. Because this limit is well understood theoretically, we can compare the numerical results with the values predicted by the conformal field theory, which serves as a benchmark for future investigations. From this study, we find that the designer model discussed above indeed dramatically suppresses the finite-size effect, and produces much more accurate scaling function of the conductivity, requiring smaller system sizes in comparison with that in the conventional models. In addition, this free-fermion simulation also provides an estimation about the low-bound of system sizes needed for studying the scaling of conductivity at the GNY QCP.

At the tree level, i.e., free Dirac fermions, the $b_1$ term in Eq.~\eqref{eq:eq10} vanishes and the scaling law can be formulated as 
\begin{eqnarray}
\langle J_{x}(\omega_n)J_{x}(-\omega_n)\rangle = -&&\sigma_{\infty}\omega_n-b_{2}\frac{T^3}{\omega^2_n} + \cdots. \label{eq:eq11}
\\
&&\text{at} \;\; \omega_n \gg T \nonumber
\end{eqnarray}
Further more, for free Dirac fermions in $(2+1)$D, it is known that the $\sigma_{\infty}=1/16$ per each fermion species~\cite{Katz2014}. Because our model contains four Dirac cones (two valleys and two spins), we will expect $\sigma_{\infty} = 1/4$ in our model.
The computed current-current correlations $\langle J_{x}(\omega_n)J_{x}(-\omega_n)\rangle$ at the bare Dirac fermion level are shown in Fig.~\ref{fig:fig5}, where the results in the main panel are for $L=50$ system and temperature $T=1/200$. For comparison, here we show the results for both the designer model (with $t_1=1$ and $t_2=1/27$) and the conventional model with ($t_1=1$ and $t_2=0$).  A fit with the form $y=-0.246x+1.097$ is also shown in figure and the corresponding $\sigma_{\infty}=0.246$ (the constant $1.097$ which does not appeared in Eq.~\eqref{eq:eq11} comes from the contribution of background of $\langle J_{x}(\omega_n)J_{x}(-\omega_n) \rangle$ at $T \rightarrow 0, \omega_n \rightarrow 0$). It is clear that the Dirac fermion in our designer model give rise to better results compared with the model with only $t_1$, and this is due to the fact that our extended linear dispersion in the BZ successfully reduced the finite energy cut-off due to the deviation of dispersion away from linear.

We carried out the same computation for a few different system sizes $L$ (while keeping $T=1/200$), and the fitted $\sigma_{\infty}$ is shown in the inset of Fig.~\ref{fig:fig5}). The dashed line in the inset indicates the exact value $\sigma_{\infty}=1/4$ at $L\to\infty$ and it is clear that as $L$ is increasing, the $\sigma_{\infty}$ obtained from the fitting are approaching $1/4$. More importantly, the (red) data points from the designer model approaches the exact value much faster than that of the conventional model (blue), and the only difference between these two model is that a small amount of next-nearest-neighbor hopping $t_2=1/27$ is introduced in the designer model. Without $t_2$, $\sigma_{\infty}$ deviates from the exact value by about $10\%$, even for the largest system size $L=50$, while the designers model has an error bar around $1\%$ at the same system size. If we compare different system sizes, the designers model with $L=30$ produces more accurate $\sigma_{\infty}$ than the conventional model at $L=50$, which implies that the designer model allows us to access quantum criticality with the same or even better accuracy at only about $1/4$ of the system size, which is consistent with the fact mentioned early on that the designer model has a larger linear-dispersion region in the BZ, about 3 times larger than the model without $t_2$.

The calculation of the conductivity at the $N_f=8$ GNY chiral Ising QCP is currently undergoing, and will be reported in a separated manuscript.


\section{conclusion}
\label{sec:iv}
Over the years, controlled study of fermionic QCPs is the question that have been haunting the minds of physicists in the field of strongly correlated systems. Although extensive efforts have been devoted and great progress has been achieved, even greater challenges are still lying in front of the community. Among those, controlled analytical calculation of the critical properties and numerical approaches such that the computational complexity ($O(\beta N^{3})$ for fermionic QMC for example) can be overcome and the thermodynamic limit can be safely reached, are the important steps towards the final solution of this challenging problem. For this objective, our efforts here focuses on improving the situation at the numeric front, utilizing new techniques in model design guided by insights gained from the field theory studies. 

The designer Hamiltonian that we have engineered realizes a lattice model of the $(2+1)$d $N_f=8$ Gross-Neveu-Yukawa (GNY) chiral Ising transition of Dirac fermions coupled to critical bosonic modes. By extending the linear-dispersion region in the BZ, and engineer the fermion and boson dispersions in order to achieve identical velocities at the bare level, we find, via QMC simulations assisted with the SLMC of the non-local bosonic configurational moves, that the strongly-coupled Dirac fermions and critical Ising bosons still acquire the almost same velocity at the QCP. In other words, the designer model place the starting point of the RG flow close to the nontrivial fixed point, and consequently minimizes the finite size effects. 

With these advances in model design and algorithm development, we could acquire robust critical exponents of GNY chiral Ising transition with only moderately large system size upto $L=16$, avoiding the heavy computational burden for pushing to larger sizes. In fact, we are able to obtain the complete set of critical exponents $\{ 1/\nu_{\text{GN}}, \ \eta_{\phi}, \ \eta_{\psi}, \ \omega\}$ than previous works (including that of ourselves~\cite{Dynamicalgeneration}) and therefore provide the much needed benchmark results for the further developments at the analytical front such as controlled RG calcuation and conformal bootstrap analysis. We also calculated the conductivity of the Dirac fermions and find its finite frequency scaling behavior consistent with conformal field theory prediction. Our approach provides the promising direction towards the eventual controlled study of fermionic QCPs, and the same concept and principles can be easily generalized to other fermionic QCPs as well, such as the GNY chiral XY~\cite{YDLiao2019} and Heisenberg~\cite{Fourloopcritical,TCLang2018} and the fermion surface coupled to the critical bosons~\cite{ZiHongLiuSqu2018}, topological orders~\cite{ChuangChen2019,Gazit2019} and Yukawa-SYK system~\cite{GPP2020}. Efforts along this line of thinking may help shed new light on key open questions in various fermionic QCPs and may eventually lead to their final solutions.

\section*{Acknowledgement}
We thank Michael Scherer, Lukas Jannsen, Joseph Maciejko, Andreas L\"auchli, Thomas Lang, Stefan Wessel, Fakher Assaad and Xi Dai for insightful discussions on both the fermionic GNY transitions and QMC methodologies, William Witczak-Krempa for enligthening conversations on the finite frequency conductivity at QCPs, Ning Su and Junchen Rong for explanations on the conformal bootstrap estimation on the fermion QCPs and Da-Chuan Lu and Yang Qi and Yi-Zhuang You for the discussion on the calculation of conductivity of Dirac fermion at finite frequency with the lattice cut-off considered. YZL, WW and ZYM acknowledge the supports from the Ministry of Science and Technology of China through the National Key Research and Development Program (Grant No. 2016YFA0300502), the Strategic Priority Research Program of the Chinese Academy of Sciences (Grant No. XDB28000000), the National Science Foundation of China (Grant Nos. 11421092,11574359,11674370) and Research Grants Council of Hong Kong Special Administrative Region of China through 17303019.  K.S. acknowledges support from the National Science Foundation under Grant No. EFRI-1741618. We thank the Center for Quantum Simulation Sciences in the Institute of Physics, Chinese Academy of Sciences, the Computational Initiative at the Faculty of Science at the University of Hong Kong and the National Supercomputer Center in Tianjin and the National Supercomputer Center in Guangzhou for their technical support and generous allocation of CPU time.
	
\begin{appendix}
\section{Velocity estimations}
\label{app:appA}

In this section, we provide analytic calculations of the velocities of free fermion and boson of our model in Eq.~\eqref{eq:eq1}.

Through Legendre transformation of $L_{\text{Fermion}}$, the fermionc $H_{\text{Fermion}}$ can be written as
\begin{equation}
H_{\text{Fermion}}=\sum_{(ij),\sigma}-t_{ij}\text{e}^{-i\sigma\theta}c_{i,\sigma}^{\dagger}c_{j,\sigma}+h.c.
\end{equation}
In the momentum space, through Fourier transformation, we obtain 
\begin{eqnarray}
\begin{split}
&H_{\text{Fermion}}=-2t_{1}\begin{pmatrix}
0& A+B \\
A^{*}+B^{*}& 0\\
\end{pmatrix}
-\frac{2t_{1}}{27}\begin{pmatrix}
0& C+D \\
C^{*}+D^{*}& 0\\
\end{pmatrix}\\
\text{where} \\
&A=\cos[\frac{\text{k}_x}{2}+\frac{\text{k}_y}{2}]\exp[i\frac{\pi}{4}] \\
&B=\cos[\frac{\text{k}_x}{2}-\frac{\text{k}_y}{2}]\exp[-i\frac{\pi}{4}]\\
&C=\cos[\frac{\text{3k}_x}{2}+\frac{\text{3k}_y}{2}]\exp[i\frac{\pi}{4}] \\
&D=\cos[\frac{\text{3k}_x}{2}-\frac{\text{3k}_y}{2}]\exp[-i\frac{\pi}{4}]
\end{split}
\label{eq:eqA1}
\end{eqnarray}
From the Eq.~\eqref{eq:eqA1}, one sees the Dirac points are located in the ($\pi$,0) and (0,$\pi$) in BZ which we then Taylor expand in the neighborhood of them,
\begin{eqnarray}
\begin{split}
H_{\text{Fermion}}=\frac{8}{9}\sqrt{2}t_{1}\begin{pmatrix}
0& -\text{k}_{x}-i\text{k}_{y}\\
-\text{k}_{x}+i\text{k}_{y}& 0\\
\end{pmatrix}
\end{split}
\end{eqnarray}

So the free fermion velocity $v_{\text{F}}=\frac{8}{9}\sqrt{2}t_{1}$ is deduced.

For the free boson which does not include the term of $\phi^4$, considering the following form of bosonic action,
\begin{eqnarray}
\begin{split}
S=\int dt\sum_{i}\frac{1}{4}(\partial_{t}\phi_{i})^2-\sum_{i,j}J_{ij}(\phi_{i}-\phi_{j})^2-\sum_{i}m\phi^2
\end{split}
\end{eqnarray}
where $J_1$, $J_2$, $J_3$ and $J_4$ are given in Fig.~\ref{fig:fig1} (c).

Again going to the momentum space via Fourier transformation, we obtain
\begin{eqnarray}
\begin{split}
S=&\sum_{\omega}\sum_{\text{q}}[\frac{1}{4}\omega^{2}-2J_{1}[(2-\cos[\text{q}_x]-\cos[\text{q}_y])\\
  &-\frac{1}{8}(2-\cos[\text{q}_x]-\cos[\text{q}_y])\\
  &+\frac{1}{63}(2-\cos[\text{q}_x]-\cos[\text{q}_y])\\
  &-\frac{1}{896}(2-\cos[\text{q}_x]-\cos[\text{q}_y])]-m]\phi_{\text{q},w}^2  
\end{split}
\end{eqnarray}

The Eular-Lagrange's equation require $S=0$ which provide the stationary solution for bosonic field $\phi$
\begin{eqnarray}
\begin{split}
\frac{1}{4}\omega^2 &=2J_{1}[(2-\cos[\text{q}_x]-\cos[\text{q}_y])-\frac{1}{8}(2-\cos[\text{q}_x]-\cos[\text{q}_y]+\\
&\frac{1}{63}(2-\cos[\text{q}_x]-\cos[\text{q}_y]))-\\
&\frac{1}{896}(2-\cos[\text{q}_x]-\cos[\text{q}_y])]+m
\end{split}
\end{eqnarray}
close to $\mathbf{Q}=(0,0)$, the dispersion is 
\begin{eqnarray}
\begin{split}
\omega &=\pm2\sqrt{\frac{5J_{1}}{8}(\text{q}_{x}^2+\text{q}_{y}^2)+m}
\end{split}
\end{eqnarray}

So the velocity of boson is $v_{b}=2\sqrt{5J_{1}/8}$. 

The condition $v_{F}=v_{b}$ require the relation of $J_{1}=0.6320t_{1}^{2}$.

\section{projective QMC and SLMC}
\label{app:appB}

\begingroup
\setlength{\tabcolsep}{6pt} 
\renewcommand{\arraystretch}{1.5} 
\begin{table*}[htp!]
	\centering
	\begin{tabular}{c c c c c}
		\hline\hline
		\ $J^{\text{eff}}_k$ & $J^{\text{eff}}_1$ & $J^{\text{eff}}_2$ & $J^{\text{eff}}_3$ & \\
		\hline
		-0.0143045485581 & -0.205175906625 &-0.0539032527282  & -0.111523376999  & \\
		\hline\hline
	\end{tabular}
	\caption{Fitted values of $J^{\text{eff}}_{k}$,   $J^{\text{eff}}_{1}$, $J^{\text{eff}}_{2}$ and $J^{\text{eff}}_{3}$ when $m=-0.06$ for $L=6$ system at $\Theta=2L$.}
	\label{tab:tab2}
\end{table*}
\endgroup

As mention in the main text, we employ the projective QMC~\cite{Dynamicalgeneration} to solve the model in Eq.~\eqref{eq:eq1}. In the QMC simulation, the partition function is written as
\begin{eqnarray}
	\begin{split}
		Z=&\langle \Phi_{T} | \text{exp}(-2 \Theta H) | \Phi_{T} \rangle \\
		=&\sum_{\left\{\phi\right\}}{W}_{\text{Boson}}\prod_{\sigma=\uparrow,\downarrow}\text{det}(P_{\sigma}^\dagger B_{m}^{\sigma}\cdots B_{1}^{\sigma}P_{\sigma}),
		\label{eq:eqB1}
	\end{split}
\end{eqnarray} 
where $\Theta$ plays the role of inverse temperature and $\phi_{\tau,i}$ is the bosonic field at imaginary time $\tau$ and site $i$. For the bosonic part of the configurational weight, the weight $W_\text{Boson}$ is of the form
\begin{eqnarray}
	\begin{split}
		W_{\text{Boson}}=\exp[-\Delta\tau(&\sum_{i,\tau}\frac{1}{4\Delta\tau^2}(\phi_{i,\tau+1}-\phi_{i,\tau})^2 + \\
		&\sum_{i,j}\sum_{\tau}V_{ij,\tau})],
	\end{split}
\end{eqnarray}
where $V_{ij,\tau}=J_{ij}(\phi_{i,\tau}-\phi_{j,\tau})^2+m\phi_{i,\tau}^2$

For the fermionic determinant, $|\Phi_{T}\rangle$ is the trial wave function which is comprised of the eigenvectors of the free feremion Hamiltonian for the occupied states, represented by a $P_{\sigma}$ matrix with size $N\times 2N$ where $N$ is the number of electrons at the half-filling. $B_{\tau}^{\sigma}$ matrix at time slice $\tau$ is then

\begin{eqnarray}
	\begin{split}
		B_\tau^{\sigma}&=\exp(-\Delta\tau(H_{\text{Fermion}}+H_{\text{Coupling}}))\\
		&=\exp(-\Delta\tau H_{\text{Fermion}})\exp(-\Delta\tau H_{\text{Coupling}}).
	\end{split}
\end{eqnarray}

Let's now compute the ratio for the fermion determinant for the QMC update. Introduce the notation
$B(\tau_{1},{\tau_{2}})=B_{\tau_1}^{\sigma}\cdots B_{\tau_{2}+1}^{\sigma}$,  $B^<=P^{\dagger}_{\sigma}B(2\Theta,\tau)$ and $B^>=B(\tau,0)P_{\sigma}$, one has
\begin{equation}
	W_{\text{Fermion}}=\det[B^< B^>], 
\end{equation}
where the Green's function is
\begin{equation}
	1-G(\tau)=B^>(B^<B^>)^{-1}B^<
\end{equation}

The local update of determinant QMC suffers from the critical slowing-down close to the QCP, to overcome this problem, we make use of the recently developed self-learning Monte Carlo (SLMC)~\cite{SLMC2016,SLMC2017} scheme to perform more efficient Monte Carlo moves in the configurational space. The core idea of SLMC is to obtain the low-energy effective bosonic Hamiltonian of the problem at hand, and this can be achieved by training an effective model from the configurations saved in a small lattice obtained from the conventional determinant QMC simulation. In our case, the Lagrangian of effective model is of the following form
\begin{equation}
\label{eq:lagrangian}
L=L_{\text{Boson}}+L_{\text{eff}}
\end{equation}
with
\begin{eqnarray}
	\begin{split}
		&L_{\text{Boson}}=\sum_{p}\left[\frac{1}{4}\big(\frac{\partial\phi_p}{\partial \tau}\big)^2+m\phi^2_p+\phi^4_p \right]+\sum_{(p,q)}J_{pq}(\phi_p-\phi_q)^2,\\ 
		&L_{\text{eff}}=J^{\text{eff}}_{k}\sum\limits_{i}\lambda_{i}\phi_{i\tau}+\sum\limits_{(ij,n)}J^{\text{eff}}_{n}\phi_{i\tau}\phi_{j\tau}
	\end{split}
\end{eqnarray}		
The spirit of the learning process is to replace the effect of $L_{\text{Fermion}}$ and $L_{\text{Coupling}}$ with  $L_{\text{eff}}$. $\lambda_{i}$ take value of $\pm1$ which has opposite signs in two bosonic sublattices.  $J_{k}^{\text{eff}}$ and $J_{n}^{\text{eff}}$-s (the $n$th-nearest interaction between the bosons) are the trained effective couplings which is shown in Tab~\ref{tab:tab2} from a $L=6$ systems at $m=-0.06$ and $\Theta=2L$. With the trained $L_\text{eff}$ and the bare boson $L_\text{Boson}$, one can then use the Lagrangian in Eq.~\eqref{eq:lagrangian} to guide the Monte Carlo move of the boson fields and only after substantially many such updates, evalue the fermion determinant in Eq.~\eqref{eq:eqB1} to respect the detailed balance condition of the original model in Eq.~\eqref{eq:eq1}. For more detailed description of SLMC, the assiduous readers are referred to the Refs.~\cite{SLMC2016,SLMC2017,XYXu2019}.

One more obstacle particularly associated with the current problem is that for training the effective model, the calculation of the fermion determinant is necessary, but the $\det[B^<B^>]$ can be very large (as large as $\text{e}^{2000}$) in our model. To overcome such numerical difficulty, we note that the real important quantity we need to calculate is $\log(\det[B^<B^>])$ in the SLMC scheme~\cite{SLMC2017}. If the $\det[B^{<}B^{>}]$ is divided into product of several parts (such as $T=A_{1}A_{2}A_{3}$), then the large value in the determinant can be breaked into sum several small parts (eg. $\log\det[T]=\log\det[A_{1}]+\log\det[A_{2}]+\log\det[A_{3}]$), where each part is small enough to be handled numerically, i.e., 
\begin{eqnarray}
	\det[P^\dagger B_s(2\Theta,0)P]&=\det[P^\dagger B^{n}B^{n-1}\cdots B^{2}B^{1}P],
\end{eqnarray}
the UDV decomposed can be used in the model
\begin{equation} 
	B^>=U^>DV
\end{equation}
where $U^>$ is a rectangular matrix $2N\times N$ and $D$ contains the $N$ eigenvalues of scale from large to small and $V$ is an upper unit triangular matrix $N \times N$. In this way, we can obtain
\begin{eqnarray}
	\begin{split}
		&\det[P^+B^{n}B^{n-1}\cdots B^{2}B^{1}P]\\
		&=\det[P^+B^{n}B^{n-1}\cdots B^{2}B^{1}P]\\
		&=\det[P^+B^{n}B^{n-1}\cdots B^{2}U^{>}_{1}D^{1}V^{1}]\\
		&=\det[P^+B^{n}B^{n-1}\cdots U^{>}_{2}D^{2}V^{2}D^{1}V^{1}]\\
		&\cdots\\
		&=\det[P^+U^{>}_{n}D^{n}V^{n}D^{n-1}V^{n-1}\cdots D^{2}V^{2}D^{1}V^{1}]\\
		&=\det[P^+U^{>}_{n}]\det[D^{n}D^{n-1}D^{n-2}\cdots D^{2}D^{1}]\det[V^{n}V^{n-1}\cdots V^{2}V^{1}],
	\end{split}
\end{eqnarray}
such that we can break the determine into three parts and compute the weight accordingly.

With the determiant weight obtained for small system sizes, we can train and obtain the parameters for the effective model. As the example shown in Tab.~\ref{tab:tab2}, one can see that interactions of several different distances $J^{\text{eff}}_{1,2,3}$ are all important as they have the same scale. With the cumulative update~\cite{SLMC2017} in SLMC, it turns out that the simulation with such effective model but applied to $L \ge 6$ systems, are sampled with shorter equilibrium time compared with conventional determinant QMC.

\section{Calculation of bosonic and fermionic dispersion}
\label{app:appC}
\begin{figure}[htp!]
\includegraphics[width =\columnwidth]{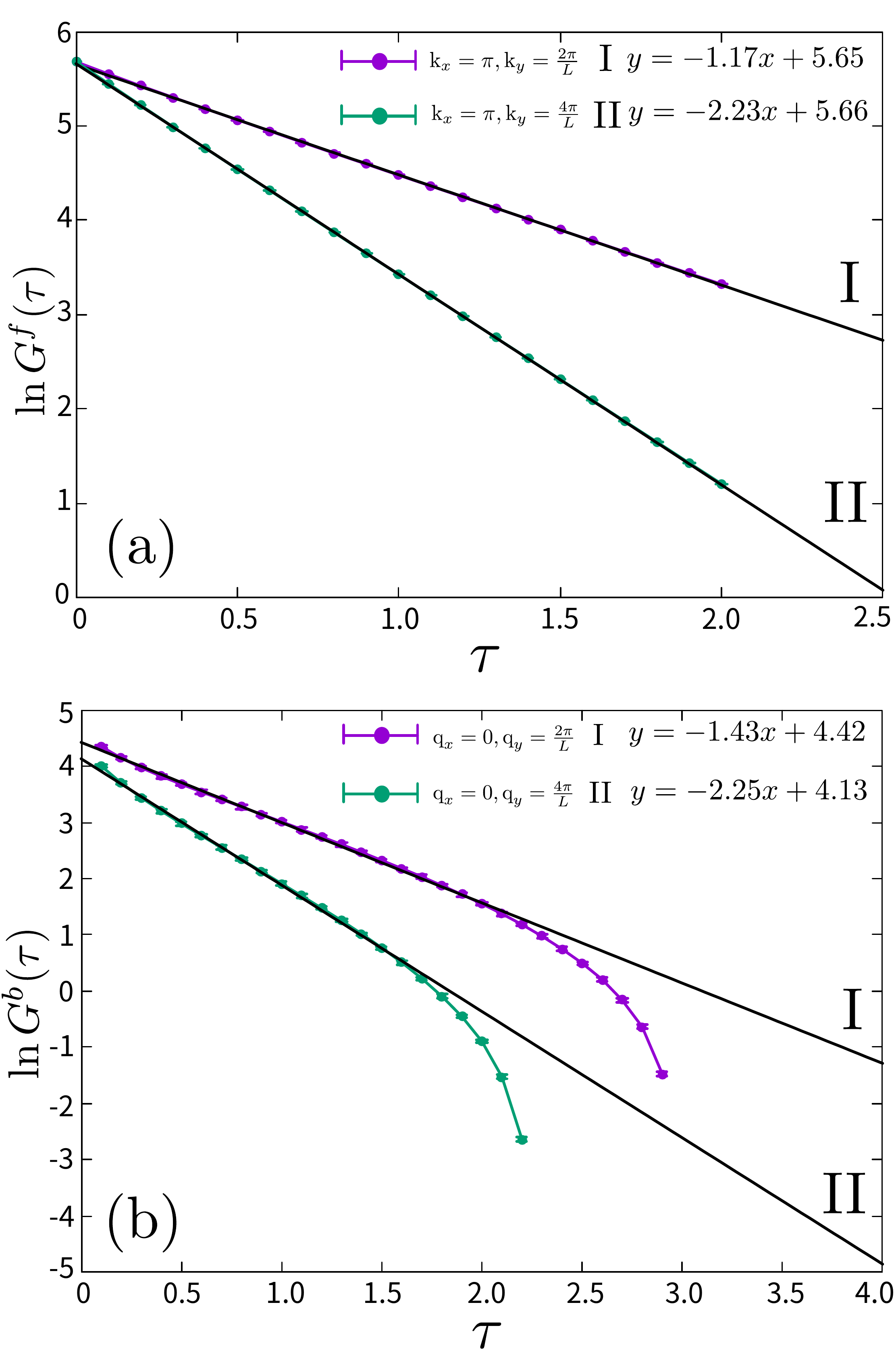}
\caption{Fermion and Boson Green's functions and their fitting according to Eq.~\eqref{eq:eqC1} from which the finite size gap $\Delta(\mathbf{k})$ at different momenta are obtained, these gaps, prepared on all the momenta along the high-symmetry-path, give rise to the dispersions $\omega(\mathbf{k})$ shown in Fig.~\ref{fig:fig1} (d) and (e) of the main text.}
	\label{fig:fig6}
\end{figure}
Here we take a closer look at how to obtain the fermionic and bosonic dispersion shown in the Fig.~\ref{fig:fig1} (d) and (e) of the main text. In the projector QMC, dynamical Green's function $G^{f}(\text{\bf{k}},\tau)$ and $G^{b}(\text{\bf{q}},\tau)$ in the the imaginary time correlation follow the Lehmann spectral representation such that at long time they satisfy the relation $G(\text{\bf{k}},\tau) \propto {e}^{-\Delta(\mathbf{k})\tau}$. So for each finite size system with $L$ momenta along the lattice axis, at each moment point $\bf{k}$, one can obtain the excitation gap, i.e. the dispersion, from the following fitting scheme,
\begin{equation}
\ln G(\mathbf{k},\tau)=-\Delta(\mathbf{k}) \tau + C,
\label{eq:eqC1}
\end{equation}
with $C$ a fitting parameter. As an example, here we show the data in Fig.~\ref{fig:fig6} (a) and (b) for fermion and boson Green's function at the closest (I) and next-closest (II) momenta away from their gapless points for $L=14$ system. For the fermionic part (shown in Fig.~\ref{fig:fig6} (a)), the obtained gap $\Delta(\mathbf{k}=(\pi,\frac{2\pi}{L}))=1.17(1)$ and $\Delta(\mathbf{k}=(\pi,\frac{4\pi}{L}))=2.23(1)$, are the values shown in the Fig.~\ref{fig:fig1} (d) which are close to the the free Dirac dispersion of our lattice model. For the bosonic part (shown in Fig.~\ref{fig:fig6} (b)), the finite size effect is a bit stronger here and at longer imaginary time, the data exhibit faster decay (at such $\tau$ the $G^{b}(\tau)$-s are very close to zero which can not be used for fitting). Nevertheless, one can still see the clear exponential part of the $G^{b}(\mathbf{q},\tau)$ and the fitting here gives rise to $\Delta(\mathbf{q}=(0,\frac{2\pi}{L}))=1.43(3)$ and $\Delta(\mathbf{q}=(0,\frac{4\pi}{L}))=2.25(3)$, are the values shown in the Fig.~\ref{fig:fig1} (e). We believe the deviation of $\Delta(\mathbf{k}=(0,\frac{2\pi}{L}))=1.43(3)$ from the linear bosonic critical dispersion is a finite size effect and this will be overcome by similar analysis but with even larger $L$, which we leave for future works.    


\end{appendix}

\bibliographystyle{apsrev4-1}
\bibliography{chiral}

\end{document}